\documentclass{article}
\usepackage{graphicx}
\usepackage{cite}
\usepackage{geometry}
\def\lsim{\raise0.3ex\hbox{$<$\kern-0.75em\raise-1.1ex\hbox{$\sim$}}} 
\def\gsim{\raise0.3ex\hbox{$>$\kern-0.75em\raise-1.1ex\hbox{$\sim$}}} 
 
\title{ The QCD transition 
temperature: results with physical masses in the continuum limit} 
\author{ 
Y. Aoki$^{a}$, Z. Fodor$^{a,b}$, 
S.D.~Katz$^{a,b}$
and K.K. Szab\'o$^{a}$\\ 
$^a$Department of Physics, University of Wuppertal, Gauss 20, 
D-42119, Germany.\\ 
$^b$Institute for Theoretical Physics, E\"otv\"os University, 
P\'azm\'any 1, H-1117 Budapest, Hungary.\\ 
} 

\begin{document} 
\maketitle
{\bf

The transition temperature ($T_c$) of QCD is determined by Symanzik
improved gauge and stout-link improved staggered fermionic lattice
simulations. We use physical masses both for the light quarks ($m_{ud}$)
and for the strange quark ($m_s$). Four sets of lattice spacings
($N_t$=4,6,8 and 10) were used to carry out a continuum extrapolation. It
turned out that only $N_t$=6,8 and 10 can be used for a controlled
extrapolation, $N_t$=4 is out of the scaling region.  Since the QCD
transition is a non-singular cross-over 
there is no unique $T_c$. Thus,
different observables lead to different numerical $T_c$ values even in the
continuum and thermodynamic limit. The peak of the renormalized chiral
susceptibility 
predicts $T_c$=151(3)(3)~MeV, wheres $T_c$-s based on the
strange quark number susceptibility and Polyakov loops result in 
24(4)~MeV and 25(4)~MeV larger values, respectively.  
Another consequence of
the cross-over is the non-vanishing width of the peaks even in the
thermodynamic limit, which we also determine. These numbers are attempted
to be the full result for the $T$$\neq$0 transition, though other lattice
fermion formulations (e.g.  Wilson) are needed to cross-check them.

}

\section{Introduction} 

The $T \neq 0$ QCD transition plays an important role in the physics of 
the early Universe and of heavy ion collisions (most recently at RHIC at 
BNL; future plans exist for the LHC at CERN and FAIR at 
GSI). In this letter we study the absolute scales of the 
transition at vanishing chemical potential ($\mu$=0), which is of direct 
relevance for the early universe ($\mu$ is negligible there) and for 
present heavy ion collisions (at RHIC $\mu$$\lsim$40~MeV, which is 
far less than the 
typical hadronic scale). The transition is known to be a cross-over 
\cite{Aoki:2006we} (at least using staggered fermions, for a discussion about 
the fourth-root trick, which is usually applied see e.g. 
\cite{Bernard:2006ee} and references therein).

There are several results in the literature for $T_c$ using both staggered 
and Wilson fermions \cite{Karsch:2000kv,Bernard:1997an,AliKhan:2000iz,Bornyakov:2004ii,Bernard:2004je,Cheng:2006qk}. 
Note however, that these results have 
typically four serious limitations. a) The first one is related to the 
unphysical spectrum. b) Another question is how to extrapolate to 
vanishing lattice spacing (a$\rightarrow$0), 
approaching the continuum limit. c) The third 
problem is how to set the absolute scale for a question, which needs an 
answer of a few percent accuracy. d) Finally the fourth problem is 
related to some implicit assumptions about a a real singularity, thus 
ignoring the analytic cross-over feature of the finite temperature QCD 
transition.

Our goal is to eliminate all these limitations and give the full answer.

{\it ad a.} All previous calculations were carried out with unphysical 
spectra. On the one hand, results with Wilson fermions were obtained with 
pion masses $m_\pi\gsim$560~MeV when approaching the thermodynamical limit
(since lattice QCD can give only dimensionless combinations, it is more 
precise to say that $m_\pi$/$m_\rho$$\gsim$0.7, where $m_\rho$ is the mass
of the rho meson). The transition is 
related to the spontaneous breaking of the chiral symmetry (which is 
driven by the pion sector) and the three physical pions have masses 
smaller  than the transition temperature, thus the numerical value of $T_c$  
could be sensitive to the unphysical spectrum. Though at $T$=0 chiral 
perturbation theory provides a technique to extrapolate to physical 
$m_\pi$, no such controllable method exists around $T_c$. On the other 
hand, results with staggered fermions suffer from taste 
violation. There is one lightest pion state and a large (usually several 
100~MeV) unphysical mass splitting between this lightest state and the 
higher lying other pion states. This mass splitting results in an 
unphysical spectrum. The artificial pion mass splitting disappears only in 
the continuum limit. For some choices of the actions the restoration of 
the proper spectrum happenes only at very small lattice spacings, whereas 
for other actions somewhat larger lattice spacings are already satisfactory.

Our solution for problem (a) is threefold. First of all, we use physical 
quark masses in our staggered analysis
(or equivalently we fix $m_K$/$f_K$ and $m_K$/$m_\pi$ to their 
experimental values, here $m_K$ and $f_K$ are the mass and decay constant
of the kaon). Secondly, 
our choice of action (stout link improved fermions) leads to smaller pion 
mass splittings than other choices (see Fig. 1 of Ref.
\cite{Aoki:2005vt}). The 
third ingredient is the continuum limit extrapolation  
which removes the pion mass splitting 
completely.

{\it ad b.} Lattice QCD uses a discretized version of the Lagrangian and
approaches the continuum limit by taking smaller and smaller lattice spacings.
For lattice spacings which are smaller than some approximate limiting value the
dimensionless ratio of different physical quantities have a specific dependence
on the lattice spacing (for staggered QCD the continuum value is approached in
this region by corrections proportional to the square of the lattice spacing).
For these lattice spacings we use the expression: 
$a^2$ scaling region. Clearly, results at least three
different lattice spacings are needed to decide, whether one is already in this
scaling region or not (two points can always be fitted by $c_0$+$c_2$$a^2$,
independently of possible large higher order terms).  Only using the $a^2$
dependencies in the scaling region, is it possible to unambiguously define the
absolute scale of the system. Outside the scaling region\footnote{Note, that
outside the scaling region even a seemingly small lattice spacing dependence
can lead to an incorrect result.  An infamous example is the Naik action
\cite{Naik:1986bn} in the Stefan-Boltzmann limit: $N_t$=4 and 6 are consistent
with each other with a few \% accuracy, but since they are not in the scaling
region they are 20\% off the continuum value.} different quantities lead to
different overall scales, which lead to ambiguous values for e.g. $T_c$.

Our solution for problem (b) is straightforward. We approach the scaling 
region by using four different sets of lattice spacing, which are 
defined as the transition region on $N_t$=4,6,8 and 
10 lattices. The results show 
(not surprisingly) that the coarsest lattice with $N_t$=4 is not in the 
$a^2$ scaling region, whereas for the other three a reliable continuum limit 
extrapolation can be carried out.

{\it ad c.} An additional problem appears if we want to give 
dimensionful predictions with a few percent accuracy. 
As we already emphasized lattice QCD predicts dimensionless combinations
of physical observables. For dimensionful predictions one calculates an
experimentally known dimensionful quantity, which is used then to set
the overall scale.
In many analyses the 
overall scale is related to some quantities which strictly speaking do 
not even exist in full QCD (e.g. the mass of the rho eigenstate and the string 
tension are not well defined due to decay or string breaking). A better, 
though still not satisfactory possibility is to use quantities, which are 
well defined, but can not be measured directly in experiments. Such
a quantity is the heavy quark-antiquark potential (V), or its
characteristic distances: the 
$r_0$ or $r_1$ parameters of V \cite{Sommer:1993ce} 
($r^2 d^2 V/dr^2$=1.65 or 1, for $r_0$ or $r_1$, respectively). 
For these quantities intermediate 
lattice calculations and/or approximations are needed to connect them to 
measurements. 
These calculations are based on $\Upsilon$ spectroscopy.
This procedure leads to further, unnecessary systematic uncertainties. 

The ultimate solution is to use 
quantities, which can be measured directly in experiments and on the 
lattice. We use the decay constant of the kaon $f_K$=159.8~MeV, which
has about 1\% measurement error. 
Detailed additional 
 analyses  were done by using the mass of the $K^*(892)$ meson $m_{K^*}$, the
pion decay constant $f_\pi$ and the value of $r_0$, which all show that 
we are in the $a^2$ scaling regime and our choice of overall scale is
unambiguous.

\begin{figure}[h!]
\centerline{\includegraphics*[width=12cm]{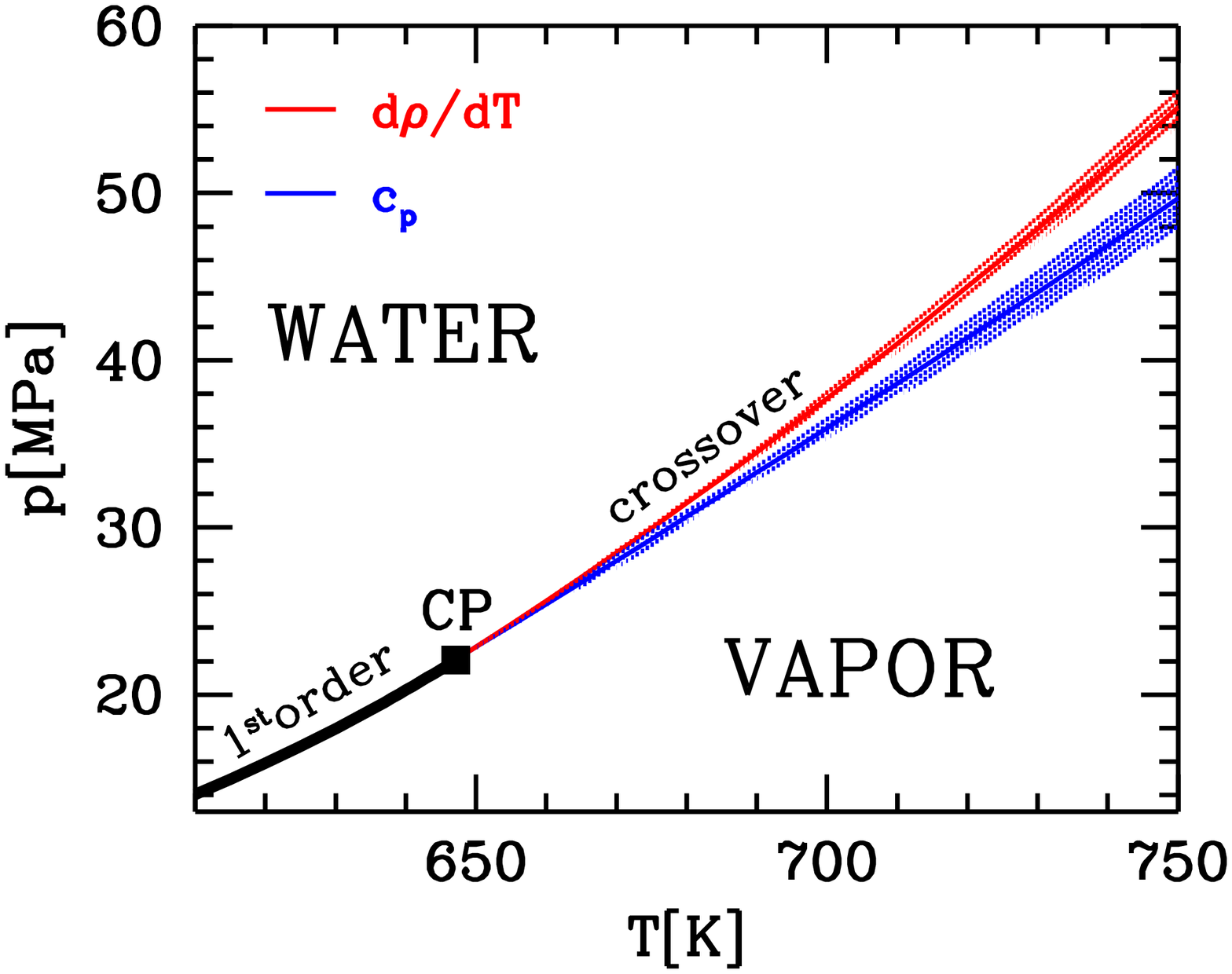}}
\caption{\label{fig:steam}
The phase diagram of water around its critical point (CP). For pressures below the critical
value ($p_c$) the transition is first order, for $p>p_c$ values there is a rapid crossover.
In the crossover region the critical temperatures defined from different quantities are not 
necessarily equal. This can be seen for the temperature derivative of the density ($d\rho/dT$) and
the specific heat ($c_p$). The bands show the experimental uncertainties (see \cite{Spang}).
}
\end{figure}

{\it ad d.} The QCD transition at non-vanishing temperatures is an 
analytic cross-over \cite{Aoki:2006we}. Since there is no singular 
temperature dependence different definitions of the transition point lead 
to different values. The most famous example for this phenomenon is the 
water-vapor transition, for which the transition temperature can be 
defined by the peaks of $d\rho/dT$ (temperature derivative of the
density) and 
$c_p$ (heat capacity at fixed pressure). For pressures ($p$) somewhat 
less than $p_c=22.064$~MPa the transition is of first order, whereas at
$p=p_c$ the 
transition is second order. In both cases the singularity guarantees that 
both definitions of the transition temperature lead to the same result. 
For $p>p_c$ the transition is a rapid cross-over, for which e.g. both
$d\rho/dT$ 
and $c_p$ show pronounced peaks as a function of the temperature, 
however these peaks are at different temperature values. Figure 
\ref{fig:steam} shows the phase diagram based on \cite{Spang}. 

Analogously, there is no unique transition temperature in QCD. Therefore, we
determine $T_c$ using the sharp changes of the temperature (T) dependence of
renormalized dimensionless quantities obtained from the chiral condensate
($\langle \bar \psi \psi \rangle$), quark number susceptibility ($n_q$) and
Polyakov loop ($P$).

The paper is organized as follows. In Section 2 we define 
our action, discuss the simulation techniques and list our simulation 
points at T=0, which will be used to carry out our continuum extrapolation
procedure unambiguously. 
Section 3 deals with the different definitions of observables, 
which are used to locate the transition point at T$\neq$0. Having located the 
transition in the lattice parameter space we make a connection to 
dimensionful physical quantities, thus determine the overall scale and 
carry out the continuum extrapolation. In 
Section 4 we conclude.

\section{Lattice action, simulations at T=0 and setting the scale}

In this letter we use a tree-level Symanzik improved gauge,
and a stout-improved staggered fermionic action (for the detailed
form of our action see eqs. (2.1-2.3) of Ref. \cite{Aoki:2005vt}). 
The stout-smearing \cite{Morningstar:2003gk} reduces the taste
violation, a lattice artefact of the staggered type of fermions.  In a previous
study we showed that this sort of smearing has the smallest taste violation
among the ones used in the literature for large scale thermodynamical
simulations. 

We have not improved the high temperature behavior of the fermion action,
however due to the order of magnitude smaller costs (compared to e.g. the p4
fat3 action) we could afford to take smaller lattice spacings ($N_t=4,6,8$ and $10$).
This 
turned out to be extremely 
 beneficial, when converting the transition temperature into
physical units.  In particular the $T=0$ simulations -which are used to do this
conversion- have very large lattice artefacts at $N_t=4$ and $6$ lattice
spacings
and can not be used for controlled continuum extrapolations. 
The high-temperature improvement is not designed to reduce these
artefacts.  

Since we want to determine the transition temperature with high precision, it
is important that the quark masses are tuned precisely to their physical
values. We have to tune the lattice quark masses along the Line of Constant
Physics (LCP) as we are approaching the continuum limit.  Using three flavor
simulations we already determined a first approximation of the LCP in
Ref.~\cite{Aoki:2005vt}.  2+1 flavor simulations with these parameters showed,
however, that the hadron mass ratios slightly differ from their physical
values.  In order to eliminate all uncertainties related to an unphysical
spectrum, we determined a new line of constant physics.  The new LCP was
defined by fixing $m_K/f_K$ and $m_K/m_\pi$ to their experimental values (c.f.
Figure 1 of \cite{Aoki:2006we}).

In order to perform the necessary renormalizations of the measured quantities
and to fix the scale in physical units we carried out $T=0$ simulations on our
new LCP (c.f. Table \ref{tab:T0}).  Six different $\beta$ values were used.
Simulations at T=0 with physical pion masses are quite expensive and in our
case unnecessary (chiral perturbation theory provides a controlled
approximation at vanishing temperature).  Thus, for each $\beta$ value we used four
different light quark masses, which resulted in pion masses somewhat larger
than the physical one (the $m_\pi$ values were approximately 250~MeV, 320~MeV,
380~MeV and 430~MeV), whereas the strange quark mass was fixed by the LCP
at each $\beta$.
The lattice sizes were chosen to satisfy the $m_\pi N_s\ge4$ condition.
However, when calculating the systematic uncertainties of meson masses and decay constants, we
have taken finite size corrections into account using continuum finite volume
chiral perturbation theory \cite{Colangelo:2005gd} (these corrections were around or less than 1\%).
We have simulated between 700 and 3000 RHMC trajectories for each point in Table \ref{tab:T0}.

Chiral extrapolation to the physical pion mass led to $m_K/f_K$ and $m_K/m_\pi$
values, which agree with the experimental numbers on the 2\% level.
(Differences resulting from various fitting forms and finite volume corrections
were included in the systematics.) This is the accuracy of our LCP.

\begin{table}
\begin{center}
\begin{tabular}{|c|c|c|c|}
\hline
$\beta$ & $m_{s}$ & $m_{ud}$ & lattice size\\
\hline
\hline
$3.330$ & 0.23847 & 0.02621 	&  $12^3\cdot24$ \\
  	&     	  & 0.04368   	&  $12^3\cdot24$\\
  	&     	  & 0.06115   	&  $12^3\cdot24$\\
  	&     	  & 0.07862   	&  $12^3\cdot24$\\
\hline                                               
$3.450$ & 0.15730 &  0.01729 	&  $16^3\cdot32$\\
  	&     	  &  0.02881 	&  $12^3\cdot28$\\
  	&    	  &  0.04033 	&  $12^3\cdot28$\\
  	&     	  &  0.05186 	&  $12^3\cdot28$\\
\hline                                               
$3.550$ & 0.10234 &   0.01312	&  $16^3\cdot32$ \\
  	&     	  &   0.01874	&  $16^3\cdot32$	  \\
  	&     	  &   0.02624	&  $12^3\cdot28$	  \\
  	&     	  &   0.03374	&  $12^3\cdot28$	  \\
\hline
 $3.670$ &  0.06331 	&  0.00928& $24^3\cdot32$  \\
 	  &  	   	&  0.01391&  $16^3\cdot32$  \\
 	  &  		&  0.01739&  $16^3\cdot32$ \\
 	  &  		&  0.02203&  $14^3\cdot32$ \\
\hline
 $3.750$ &  0.05025 	&  0.00736& $24^3\cdot32$   \\
 	  &  		&  0.01104&  $24^3\cdot32$   \\
 	  &  		&  0.01473&  $16^3\cdot32$   \\          
 	  &  		&  0.01841&  $16^3\cdot32$  \\
\hline

\end{tabular}          
\end{center}
\caption{\label{tab:T0}
Lattice parameters and sizes of our zero temperature simulations. The strange quark mass
is varied along the LCP as $\beta$ is changed. The light quark masses,
listed at each ($\beta$,$m_s$) values, correspond 
approximately to $m_\pi$ values of 250~MeV, 320~MeV,  380~MeV and 430~MeV.}
\end{table}

In order to be sure that our results are safe from ambiguous determination
of the overall scale, and to prove that we are really in the $a^2$
scaling region,  we carried out a continuum extrapolation for three
additional quantities which could be similarly good to set the scale (we normalized
them by $f_K$, for $f_K$ determination in staggered QCD see \cite{Aubin:2004fs}).  Figure \ref{fig:spect} shows the measured values of
$m_{K^*}/f_K$, $f_\pi/f_K$ and $r_0f_K$, at different lattice spacings and
their continuum extrapolation. Our three continuum predictions are in complete
agreement with the experimental results (note, that $r_0$ can not be measured
directly in experiments; in this case the original experimental input is the spectrum of the
$\Upsilon$ resonance, which was used by the MILC, HPQCD and UKQCD collaborations to calculate
$r_0$ on the lattice \cite{Aubin:2004fs,Gray:2005ur}).

It is important to emphasize
that at lattice spacings given by $N_t$=4 and 6 the overall 
scales determined by $f_K$ and
$r_0$ are differing by $\sim$20-30\%, which is most probably true for any
other staggered formulation used for thermodynamical calculations. Since
the determination of the overall scale has a $\sim$20-30\% ambiguity, the
value of $T_c$ can not be determined with the required accuracy.

For the simulations we were using the RHMC algorithm with multiple time
scales \cite{Clark:2006fx}. The time consuming parts of the computations were carried out in
single precision, however the exact
reversibility of the algorithm
was achieved. On one of our largest
lattices we have cross-checked the results with a fully double precision
calculation.

\section{T$\neq$0 simulations, transition points for different observables}

The T$\neq$0 simulations (c.f. Table \ref{tab:T}) were carried out along our LCP
(that is at physical strange and light quark masses,
which correspond to $m_K$=498~MeV and $m_\pi$=135~MeV) at four different
sets of lattice spacings ($N_t=4,6,8$ and $10$) and on three different volumes
($N_s/N_t$ was ranging between 3 and 6). We have observed moderate finite
volume effects on the smallest volumes for quantities which are supposed to depend 
strongly on light quark masses (e.g.. chiral susceptibility). To determine
the transition point we used $N_s/N_t\ge 4$, for which we did not
observe any finite volume effect. The number of RHMC trajectories were between 1500 and 8000
for each parameter set (the integrated autocorrelation time was smaller 
or around 10 for all our runs).

\begin{table}
\begin{center}
\begin{tabular}{|c|c|c|}
\hline
temporal size ($N_t$) & $\beta$ range & spatial sizes ($N_s$)\\
\hline
\hline
$4$   &	$3.20-3.50$	&	$12,16,24$	\\
$6$   &	$3.45-3.75$	&	$18,24,32$	\\
$8$   &	$3.57-3.76$	&	$24,32,40$	\\
$10$  &	$3.63-3.86$	&	$28,40,48$	\\
\hline	
\end{tabular}
\end{center}
\caption{\label{tab:T}Summary of the T$\neq$0 simulation points.}
\end{table}

We considered three quantities to locate the transition point: the chiral
susceptibility, the strange quark number
susceptibility and the Polyakov-loop. 
Since the transition at vanishing chemical potential 
is a cross-over, we expect that 
all three quantities result in different transition points (similarly
to the case of the water, c.f. Figure \ref{fig:steam}). 

\subsection*{Chiral susceptibility}

The chiral susceptibility of the light quarks ($\chi$) is defined as 
\begin{equation} 
\chi_{\bar{\psi}\psi}=\frac{T}{V}\frac{\partial^2}{\partial m_{ud}^2} \log Z= 
-\frac{\partial^2}{\partial m_{ud}^2}f, 
\end{equation}
where $f$ is the free energy density. Since both the bare quark mass and
the free energy density contain divergences, $\chi_{\bar{\psi}\psi}$ has to be renormalized
\cite{Aoki:2006we}.

The renormalized quark mass can be written as $m_{R,ud}=Z_m\cdot m_{ud}$.
If we apply a mass independent renormalization then we have
\begin{equation} 
m_{ud}^2\frac{\partial^2}{\partial
m_{ud}^2}=m_{R,ud}^2\frac{\partial^2}{\partial m_{R,ud}^2}. 
\end{equation}
The free energy has additive, quadratic divergencies. They can be removed by
subtracting the free energy at $T=0$ (this is the usual renormalization
procedure for the free energy or pressure), which leads to $f_R$. Therefore, we have the following
identity: 
\begin{equation}
m_{ud}^2\frac{\partial^2}{\partial
m_{ud}^2}\left(f(T)-f(T=0)\right)= m_{R,ud}^2\frac{\partial^2}{\partial
m_{R,ud}^2}f_R(T). 
\end{equation} 
the right hand side contains only renormalized quantities, which
can be determined by measuring the susceptibilities of the left hand side 
(for the above expression we use the shorthand notation
$m_{ud}^2 \cdot \Delta\chi_{\bar{\psi}\psi}$).
In order to obtain a dimensionless quantity
it is natural to  normalize the above quantity by $T^4$
(which minimizes the final errors). 
Alternatively, one can use 
combinations of $T$ and/or $m_\pi$ to construct dimensionless
quantities (though these conventions lead to larger errors). 
Since the transition is a cross-over (c.f. discussion d of our Introduction)
the maxima of $m_{ud}^2/m_\pi^2 \cdot \Delta\chi_{\bar{\psi}\psi}/T^2$ or
$m_{ud}^2/m_\pi^4 \cdot \Delta\chi_{\bar{\psi}\psi}$ give somewhat different values for
$T_c$. 

The upper panel of 
Figure \ref{fig:susc} shows the temperature dependence of the 
renormalized chiral susceptibility
for different temporal extensions ($N_t$=4,6,8 and 10). 
For small enough lattice spacings, thus close to the continuum limit,
these curves should coincide.  As it can be seen, the $N_t=4$
result has considerable lattice artefacts,
however the two smallest lattice
spacings ($N_t=8$ and $10$) are already consistent with each other,
suggesting that they are also consistent with the continuum limit 
extrapolation (indicated by the orange band). The curves exhibit pronounced peaks. We define the 
transition temperatures by the position of these peaks. We 
fitted a second order expression to the peak to obtain its position. 
The slight
change due to the variation of the fitting range is taken as a systematic
error. The left panel of Figure \ref{fig:tc} shows the transition 
temperatures in
physical units for different lattice spacings obtained from the
chiral susceptibility. As it can be seen
$N_t$=6,8 and 10 are already in the scaling region, thus a safe
continuum extrapolation can be carried out.
The extrapolations based on $N_t=6,8,10$ fit and
$N_t=8,10$ fit are consistent with each other. For our final result we
use the average of these two fit results (the difference between them
are added to our systematic uncertainty). 
Our T=0 simulations resulted in a $2\%$ error on the overall scale.
Our final result for the transition temperature based on the chiral
susceptibility reads:  
\begin{equation} 
T_c(\chi_{\bar{\psi}\psi})=151(3)(3) {\rm ~MeV},
\end{equation} 
where the first error comes from the T$\neq$0, the second from
the T=0 analyses.  

We use the second derivative of the chiral susceptibility ($\chi''$) at the peak position
to estimate the width of the peak ($(\Delta T_c)^2 = - \chi(T_c)/\chi''(T_c)$).
For the continuum
extrapolated width we obtained:
\begin{equation}
\Delta T_c(\chi_{\bar{\psi}\psi})=28(5)(1) {\rm ~MeV.}
\end{equation}
  
Note, that for a real phase transition (first or second order), the
peak would have a vanishing width
(in the thermodynamic limit), 
yielding a unique value for the critical
temperature. Due to the crossover nature of the transition there is no such
value, there is a range ($151 \pm 28$ MeV) where the transition phenomena
takes place. Other quantities than the chiral susceptibility could
result in transition temperatures within this range.

The MILC collaboration also reported a continuum result on the transition
temperature based on the chiral susceptibility \cite{Bernard:2004je}. Their result is 
169(12)(4)~MeV. Note, that
their lattice spacings were not as small as ours
(they used $N_t$=4,6 and 8), their aspect ratio was quite small
($N_s$/$N_t$=2), they used non-physical quark masses (their smallest
pion mass at T$\neq$0 was $\approx$220~MeV),   
the non-exact R-algorithm was applied for the simulations
and they did not use the renormalized susceptibility, but
they looked for the peak in the bare $\chi_{\bar{\psi}\psi}/T^2$.
Using $T^4$ as a normalization prescription (as we did) 
the transition temperature would decrease their $T_c$ values
by approximately $9$~MeV.
Note, that their continuum extrapolation resulted in a quite
large error. Taking into account their uncertainties
our result and their result agree on the 1-sigma level.

\subsection*{Quark number susceptibility}

For heavy-ion experiments the quark number susceptibilities are 
quite useful, since
they could be related to event-by-event fluctuations.
Our second transition temperature is obtained from the strange quark number
susceptibility,  which is defined via 
\cite{Bernard:2004je}
\begin{equation}
\frac{\chi_{s}}{T^2}=\frac{1}{TV}\left.\frac{\partial^2 \log Z}{\partial \mu_{s} ^2
}\right|_{\mu_{s}=0}, 
\end{equation}
where $\mu_s$ is the strange quark chemical potential (in
lattice units). Quark number susceptibilities have the convenient property,
that they automatically have a proper continuum limit, there is no need for
renormalization.

The middle panel of
Figure \ref{fig:susc} shows the temperature dependence of the
strange quark number susceptibility
for different temporal extensions ($N_t$=4,6,8 and 10).
For small enough lattice spacings, thus close to the continuum limit,
these curves should coincide again (our continuum limit estimate is 
indicated by the orange band).

As it can be seen, the $N_t=4$
results are quite off, however the two smallest lattice
spacings ($N_t=8$ and $10$) are already consistent with each other,
suggesting that they are also consistent with the continuum limit
extrapolation. This feature indicates, that they are closer to 
the continuum result than our statistical uncertainty.

We defined the transition temperature as the peak in the temperature
derivative of the strange quark number susceptibility, 
that is the inflection point of the
susceptibility curve. The position was determined by two independent ways,
which yielded the same result. In the first case we fitted a cubic
polynomial on the susceptibility curve, while in the second case we
determined the temperature derivative numerically from neighboring points
and fitted a quadratic
expression to the peak. The slight
change due to the variation of the fitting range is taken as a systematic
error. The middle panel of Figure \ref{fig:tc} shows the transition
temperatures in
physical units for different lattice spacings obtained from the
strange quark number susceptibility. As it can be seen
$N_t$=6,8 and 10 are already in the $a^2$ scaling region, thus a safe
continuum extrapolation can be carried out.
The extrapolations based on $N_t=6,8,10$ fit and
$N_t=8,10$ fit are consistent with each other. For our final result we
use the average of these two fit results (the difference between them
is added to our systematic uncertainty).
The continuum extrapolated value for the transition temperature 
based on the strange quark number susceptibility is
significantly higher than the one from the chiral susceptibility. The 
difference is 24(4)~MeV.  For the transition temperature in the continuum 
limit one gets:
\begin{equation}
T_c(\chi_s)=175(2)(4) {\rm ~MeV},
\end{equation}
where the first (second) error is from the T$\neq$0 (T=0) temperature
analysis (note, that due to the uncertainty of the
overall scale, the difference is more precisely determined than the
uncertainties of 
$T_c(\chi_{\bar{\psi}\psi})$
and $T_c(\chi_s)$ would suggest).
\footnote{ 
A continuum extrapolation using only the two 
coarsest lattices ($N_t=4$ and $6$) 
yielded
$T_c \sim 190$~MeV \cite{Katz:2005br}, 
where an approximate LCP was used,
if the lattice spacing is set by $r_0$.
}
Similarly to the 
chiral susceptibility analysis, the curvature at the peak can be used
to define a width for the transition. 
\begin{equation}
\Delta T_c(\chi_s)= 42(4)(1) {\rm ~MeV}.
\end{equation}

\subsection*{Polyakov loop}

In pure gauge theory the order parameter of the confinement transition is 
the Polyakov-loop: 
\begin{equation} 
P=\frac{1}{N_s^3}\sum_{\bf x} {\rm tr} 
[U_4({\bf x},0) U_4({\bf x},1) \dots U_4({\bf x},N_t-1)]. 
\end{equation} 
P acquires a non-vanishing expectation value in the deconfined phase, 
signaling the spontaneous breakdown of the Z(3) symmetry. When fermions 
are present in the system, the physical interpretation of the 
Polyakov-loop expectation value is more complicated (see e.g..
\cite{Kratochvila:2006jx}). 
However, its absolute value can be related to the quark-antiquark 
free energy at infinite separation: 
\begin{equation} 
|\langle P \rangle |^2 = 
\exp(-\Delta F_{q\bar{q}}(r\to \infty)/T). 
\end{equation} 
$\Delta F_{q\bar{q}}$ is the difference of the free energies 
of the quark-gluon plasma with and without the quark-antiquark pair.

The absolute value of the Polyakov-loop vanishes in the continuum 
limit. It needs renormalization. This can be 
done by renormalizing the free energy of the quark-antiquark pair
\cite{Kaczmarek:2002mc}. 
Note, that QCD at T$\neq$0 has only the  
ultraviolet divergencies which are already present at T=0. 
In order to remove these divergencies at a given lattice spacing
we used a simple renormalization condition \cite{Fodor:2004ft}: 
\begin{equation}
V_R(r_0)=0,
\end{equation} 
where the potential is measured at T=0 from Wilson-loops. The above
condition fixes the additive term in the potential at a given lattice
spacing. This additive term can be used at the same lattice spacings
for the potential obtained from Polyakov loops, or equivalently
it can be built in into the definition of the renormalized 
Polyakov-loop.
\begin{equation} 
|\langle P_R \rangle | = |\langle 
P \rangle | \exp(V(r_0)/(2T)), 
\end{equation}
where $V(r_0)$ is the unrenormalized potential  
obtained from Wilson-loops.

The lower panel of
Figure \ref{fig:susc} shows the temperature dependence of the
renormalized Polyakov-loops
for different temporal extensions ($N_t$=4,6,8 and 10).
The two smallest lattice
spacings ($N_t=8$ and $10$) are approximately in 1-sigma agreement
 (our continuum limit estimate is indicated by the orange band).

Similarly to the strange quark susceptibility case
we defined the transition temperature as the peak in the temperature
derivative of the Polyakov-loop, that is the inflection point of the
Polyakov-loop curve. To locate this point and determine its uncertainties
we used the same two methods,
which were used to determine $T_c(\chi_s)$.  
The right panel of Figure \ref{fig:tc} shows the transition
temperatures in
physical units for different lattice spacings obtained from the
Polyakov-loop. As it can be seen
$N_t$=6,8 and 10 are already in the scaling region, thus a safe
continuum extrapolation can be carried out. The extrapolation
and the determination of the systematic error were done as for
$T_c(\chi_s)$.
The continuum extrapolated value for the transition temperature 
based on the renormalized Polyakov-loop is
significantly higher than the one from the chiral susceptibility. The 
difference is 25(4)~MeV.  For the transition temperature in the continuum 
limit one gets:
\begin{equation}
T_c(P)=176(3)(4) {\rm ~MeV},
\end{equation}
where the first (second) error is from the T$\neq$0 (T=0) temperature
analysis (again, due to the uncertainties of the
overall scale, the difference is more precisely determined than the
uncertainties of $T_c(\chi)$ and $T_c(P)$ suggest). Similarly to the
chiral susceptibility analysis, the curvature at the peak can be used
to define a width for the transition.
\begin{equation}
\Delta T_c(P)=38(5)(1) {\rm ~MeV}.
\end{equation}

\section{Conclusions} 

We determined the transition temperature of QCD by Symanzik
improved gauge and stout-link improved staggered fermionic lattice
simulations. We used an exact simulation algorithm and 
physical masses both for the light quarks 
and for the strange quark. The parameters were 
tuned with a quite high precision,
thus at all lattice spacings the $m_K/f_k$ and $m_k/m_\pi$ ratios
were set to their experimental values with an accuracy better than 2\%.  
Four sets of lattice spacings
($N_t$=4,6,8 and 10) were used to carry out a continuum extrapolation. It
turned out that only $N_t$=6,8 and 10 can be used for a controlled
extrapolation, $N_t$=4 is out of the scaling region. Lattice spacings
obtained at the  $N_t$=6,8 and 10 transition points still result in
different values for different physical inputs, but they
  are already in the scaling region, and an $a^2$ type extrapolation
can be used. Any extrapolation merely 
based on $N_t$=4 and 6 would contain an unknown systematic error. 
We demonstrated, that our result is independent of the choice of the physical 
quantity, which is used to set the overall scale. We calculated three 
additional quantities, which would give the same dimensionful result
for $T_c$, since we reproduced their experimental values in the 
continuum limit. (These ambiguities, related to setting the scale, are
serious drawbacks of the analyses which can be found in 
the literature.)   

Since the QCD transition is a non-singular cross-over \cite{Aoki:2006we} 
there is no unique $T_c$. We illustrated this well-known phenomenon on the
water-vapor phase diagram. 
Different observables lead to different numerical $T_c$ values even in the
continuum and thermodynamic limit also in QCD. We used three observables to 
determine 
the corresponding transition temperatures. The peak of the 
renormalized chiral
susceptibility predicts $T_c$=151(3)(3)~MeV, whereas $T_c$-s based on the
strange quark number susceptibility resulted in 24(4)~MeV larger value.
Another quantity, which is related to the confining phase transition
in the large quark mass limit is the Polyakov loop. Its behavior predicted
a 25(4)~MeV larger transition temperature, than that of the chiral
susceptibility.
Another consequence of
the cross-over is the non-vanishing widths of the peaks even in the
thermodynamic limit, which we also determined. For the chiral susceptibility,
strange quark number susceptibility and Polyakov-loop we obtained widths of 
28(5)(1)~MeV, 42(4)(1)~MeV and 38(5)(1)~MeV, respectively. These numbers are attempted
to be the full result for the $T$$\neq$0 transition, though other lattice
fermion formulations (e.g.  Wilson) are needed to cross-check them.

{\bf Note.} After finishing the simulations of the present paper
and preparing our manuscript an independent study on $T_c$ 
based on large scale simulations of the  
Bielefeld-Brookhaven-Columbia-Riken group appeared on the archive
\cite{Cheng:2006qk}.
The p4fat3 action was used, which is designed to give very good results 
in the (T$\longrightarrow$$\infty$) Stefan-Boltzmann limit (their 
action is not optimized at T=0, which is needed e.g. to set the scale). 
The overall scale was set by $r_0$. 
The $T_c$ analysis based on the chiral susceptibility
peak gave in the continuum limit $T_c(\chi)$=192(7)(4)~MeV.
(The second error, 4~MeV,  estimates the uncertainty of the continuum limit 
extrapolation, which we do not use in the following, since we 
attempt to give a more reliable estimate on that.)
This result is in obvious contradiction with our continuum result from 
the same observable, which is $T_c(\chi)$=151(3)(3)~MeV. For the same
quantity (position of chiral susceptibility peak with physical quark 
masses in the continuum limit) one should obtain the same numerical result 
independently of the lattice action. Since the chance probability that 
we are faced with a statistical fluctuation and
both of the results are correct is small, we attempted to
understand the origin of the discrepancy. 
We repeated some of their simulations and analyses. 
In these cases a complete agreement
was found. In addition to their T=0 analyses we carried out
an $f_K$
determination, too. This $f_K$ was used to extend their work, 
to use an LCP
based on $f_K$  and to determine $T_c$ in physical units. 

We summarize the origin of the contradiction between our findings and theirs. 
The major part of the difference can be explained by the fact,
that the lattice spacings of  
\cite{Cheng:2006qk} are too large ($\gsim$0.20~fm), thus they are
not in the $a^2$ scaling regime, in which a justified continuum 
extrapolation could have been done. 
Setting the scale by different dimensionful quantities
should lead to the same result. However at their lattice spacings the
overall scales obtained by $r_0$ or by $f_K$ can differ by $\gsim$20\%,
and even the continuum extrapolated $r_0f_K$ 
value of these scales is about 4--5$\sigma$ away from
the value given by the literature 
\cite{Aubin:2004fs,Gray:2005ur}.
This scale ambiguity appears in $T_c$, too (though other uncertainties
of \cite{Cheng:2006qk},  e.g. coming from the determination of the 
peak-position, somewhat hide its high statistical significance). 
We used their $T_c$ values fixed by their $r_0$ scale, and in addition we
converted their peak position of the chiral susceptibility to $T_c$
setting the scale by $f_K$. (In order to ensure the possibility of 
a consistent continuum limit --independently of the actual physical
value of $r_0$-- we used for both $r_0$ and $f_K$ the results of 
\cite{Aubin:2004fs,Gray:2005ur} as Ref. \cite{Cheng:2006qk} did it for $r_0$.)
Setting the overall scale by $f_K$ predicts a much smaller $T_c$
at their lattice spacings
than doing it by $r_0$ (see left panel of Figure
\ref{fig:note}). Even after carrying out 
the continuum extrapolation the difference
does not vanish ($\sim 30$~MeV), which means that the lattice spacings 
$\gsim$0.20~fm used by
\cite{Cheng:2006qk} are not in the scaling regime.  Thus, results obtained
with their lattice spacings can not give a consistent continuum limit for $T_c$.

\begin{figure}[h!]
\centerline{\includegraphics*[width=12cm,height=15cm]{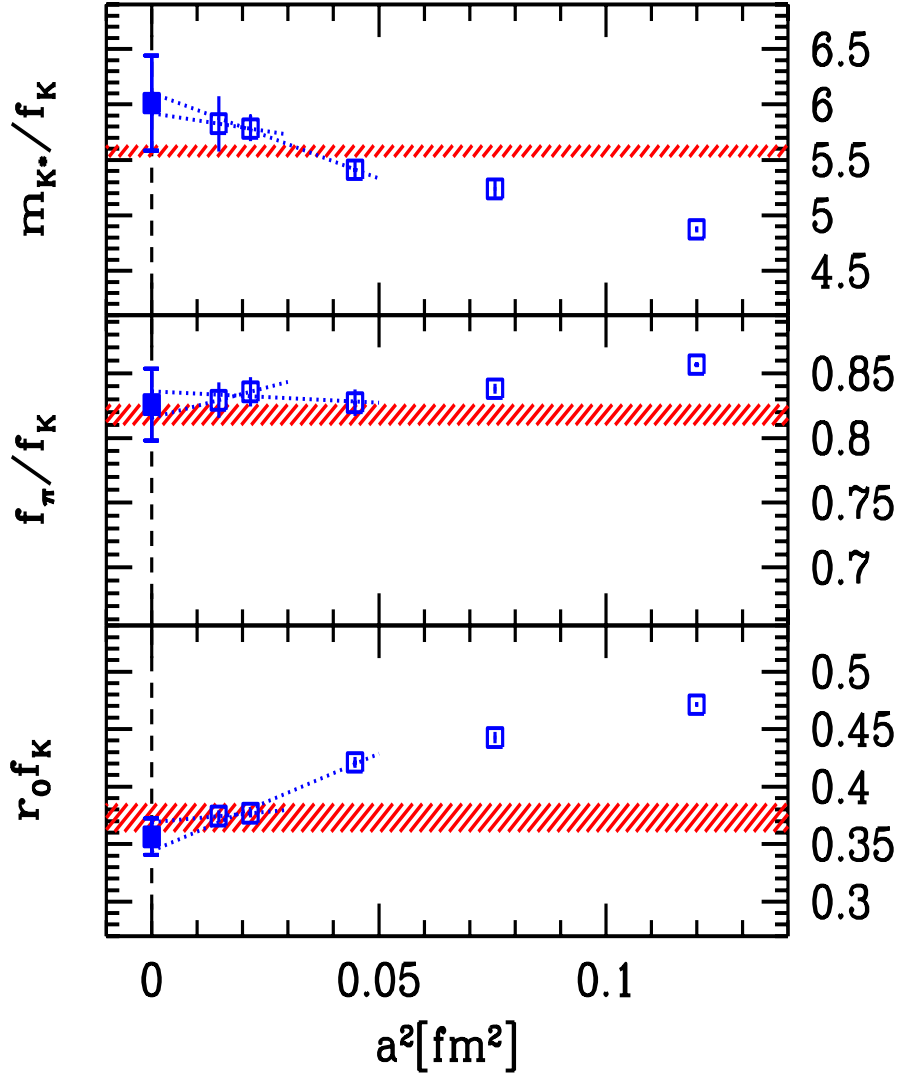}}
\caption{\label{fig:spect}
Scaling of the mass of the $K^*(892)$ meson, the pion decay constant and $r_0$ 
towards the continuum limit.
As a continuum value (filled boxes) we took the average of the continuum extrapolations obtained using our 2 and our 3
finest lattice spacings. The difference was taken as a systematic uncertainty, which is included in the shown errors.
The quantities are plotted in units of the kaon decay constant. 
In case of the upper two panels the bands indicate the physical values of the ratios and their
experimental uncertainties. For $r_0$ (lowest panel) in the absence of direct
experimental results
we compare
our value with the $r_0f_K$
obtained by the MILC, HPQCD and UKQCD collaborations 
\cite{Aubin:2004fs,Gray:2005ur}.
}
\end{figure}

\begin{figure}[h!]
\centerline{\includegraphics*[height=20cm]{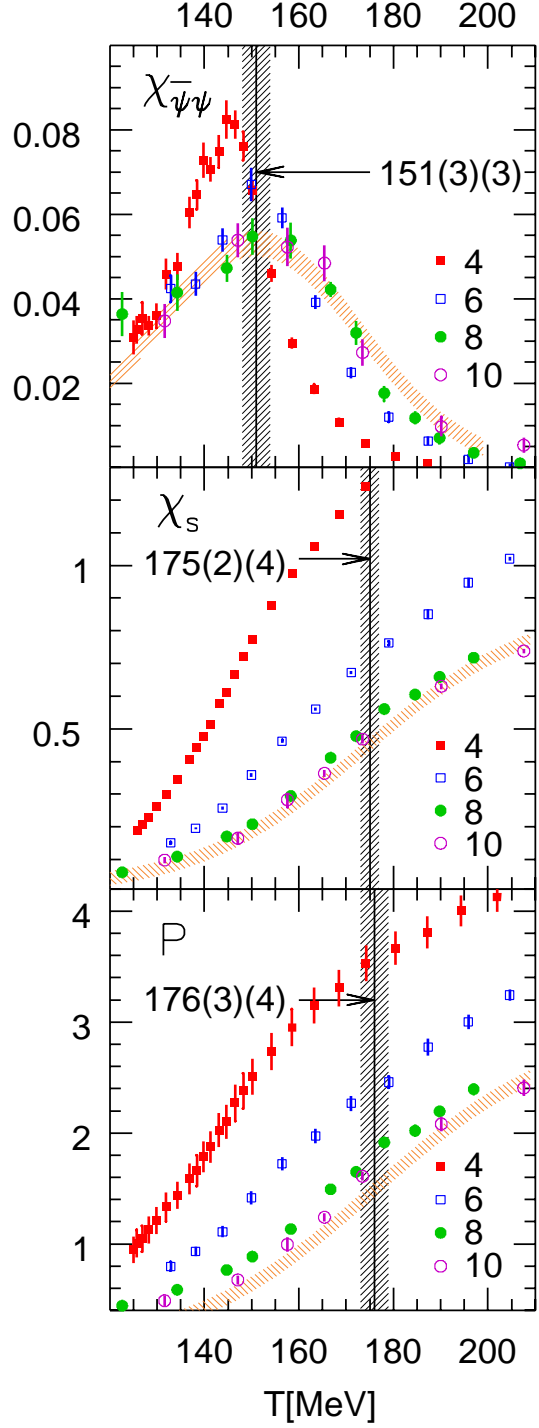}}
\caption{\label{fig:susc}
Temperature dependence of the renormalized
chiral susceptibility ($m^2\Delta \chi_{\bar{\psi}\psi}/T^4$), the strange
quark number susceptibility ($\chi_s/T^2$) 
and the renormalized Polyakov-loop ($P_R$) in the transition region. The different symbols show the results for $N_t=4,6,8$ and
$10$ lattice spacings (filled and empty boxes for $N_t=4$ and $6$, filled and open circles for $N_t=8$
and $10$).
The vertical bands indicate the corresponding critical temperatures and 
its uncertainties coming from the T$\neq$0 analyses. This error is 
given by the number in the first parenthesis, whereas the error of the 
overall scale determination is indicated by the number in the second 
parenthesis. The orange bands show our continuum limit estimates for the 
three renormalized quantities as 
a function of the temperature with their uncertainties.
}
\end{figure}

\begin{figure}[h!]
\centerline{\includegraphics*[height=6cm,bb=18 520 589 704]{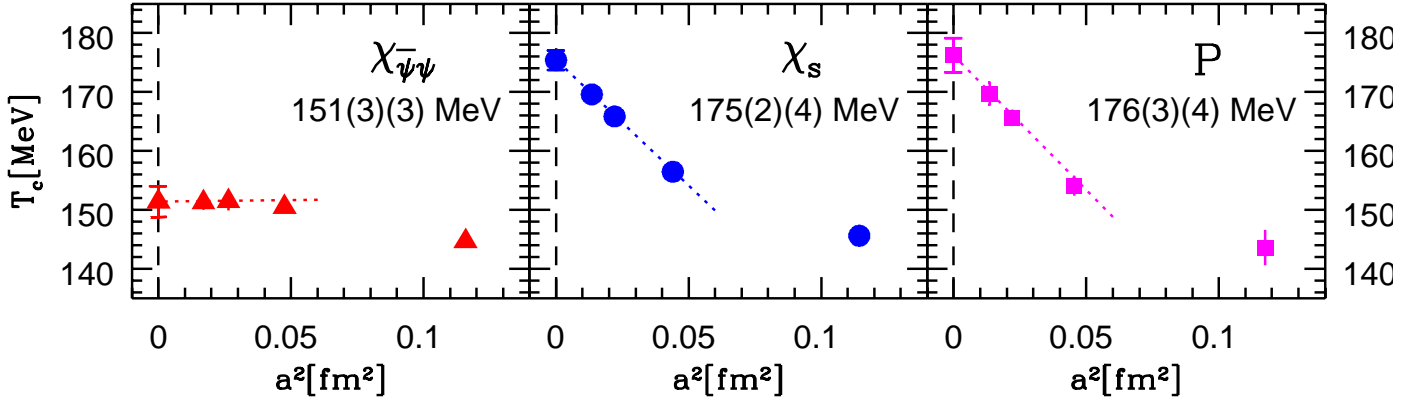}}
\caption{\label{fig:tc}
Continuum limit of the transition temperatures obtained from the renormalized chiral
susceptibility ($m^2\Delta \chi_{\bar{\psi}\psi}/T^4$), 
strange quark number susceptibility ($\chi_s/T^2$) and renormalized
Polyakov-loop ($P_R$). 
}
\end{figure}

\begin{figure}[h!]
\centerline{\includegraphics*[width=18cm,bb=90 620 492 816]{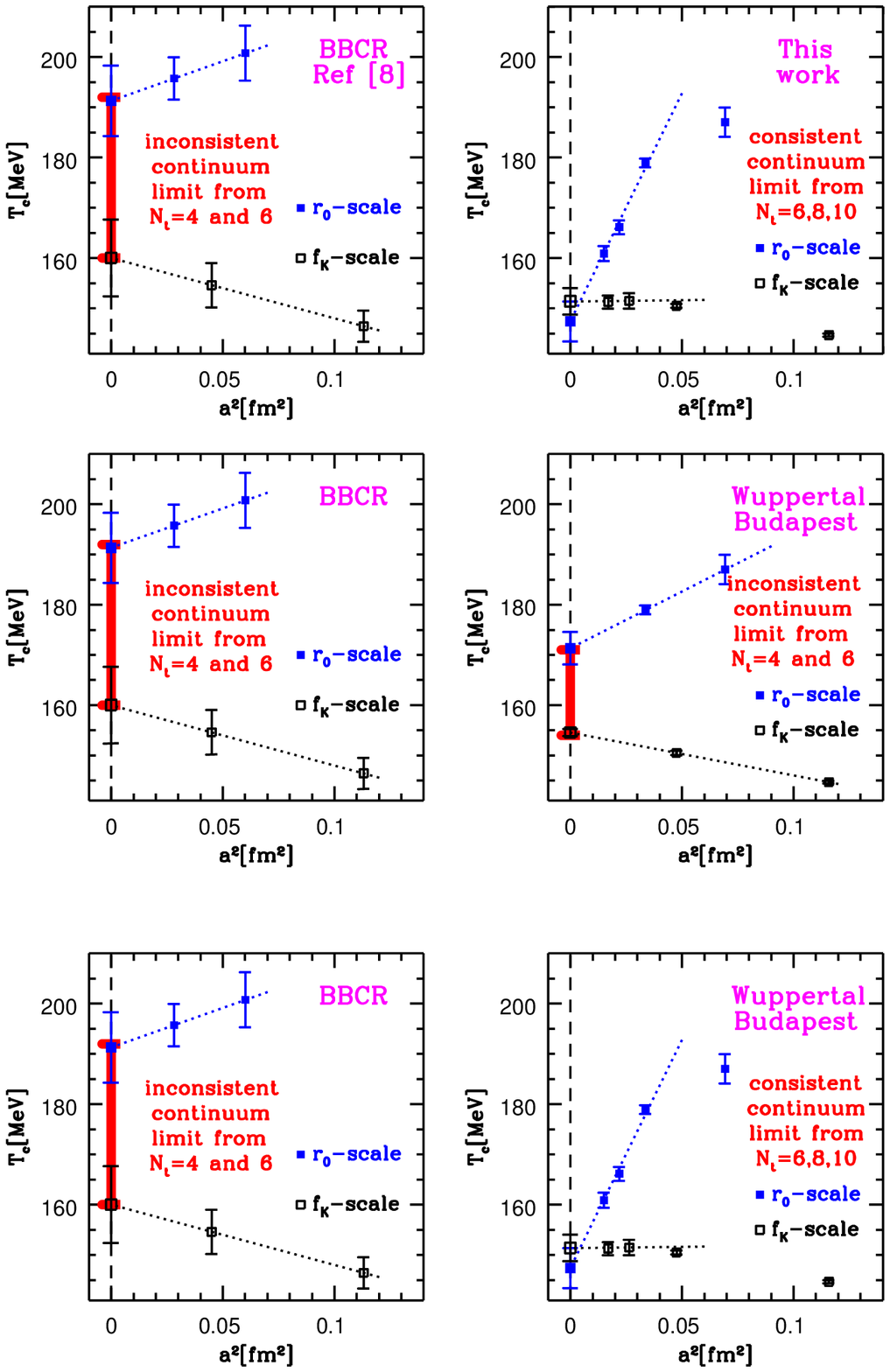}}
\caption{\label{fig:note}
Resolving the discrepancy between the critical temperature of Ref.
\cite{Cheng:2006qk} and that of the present work (see text). 
The major part of the difference can be traced
back to the unreliable continuum extrapolation of \cite{Cheng:2006qk}.  Left
panel: In Ref. \cite{Cheng:2006qk} $r_0$ was used for scale setting
(filled boxes), however using the kaon decay constant 
(empty boxes) leads to different critical temperatures even after performing 
the
continuum extrapolation.  Right panel: in our work the extrapolations based on
the finer lattices are safe, using the two different scale setting methods one
obtains consistent results.  
}
\end{figure}

In our case not only $N_t=4$ and $6$ temporal extensions were used,
but more realistic $N_t$=8 and $10$ simulations were carried out,
which led to smaller lattice spacings. These calculations
are already in the $a^2$ scaling regime and a safe continuum
extrapolation can be done. For our lattice spacing 
different scale setting methods give  consistent results. This is
shown on the right panel of Figure \ref{fig:note} (independently of the
scale setting one obtains the same $T_c$) and also justified with
high accuracy by Figure
\ref{fig:spect}, where $r_0f_K$ converges to the physical value on our finer
lattices. 
As it can be seen on the plot, using only our $N_t=4$ and 6 results would also give an inconsistent continuum limit. This
emphasizes our conclusion that lattice spacings $\gsim$ 0.20~fm can not be used for consistent continuum extrapolations. 

The second, minor part of the difference comes from the different
definitions of the critical temperatures related to the chiral
susceptibility.  We use the renormalized chiral susceptibility 
with $T^4$ normalization to obtain the
peak position, which yields $\sim 9$~MeV smaller critical temperature than the
bare susceptibility normalized by $T^2$ of Ref. \cite{Cheng:2006qk}. 

{\bf Acknowledgment:}
We thank F.~Csikor, J.~Kuti and L.~Lellouch
for useful discussions. We thank P.~Petreczky for critical
reading of the manuscript.
This research was partially supported by
OTKA Hungarian Science Grants No.\ T34980, T37615, M37071, T032501, AT049652,
by DFG German Research Grant No.\ FO 502/1-1
and by the EU Research Grant No.
RII3-CT-20040506078.
The computations were carried out on the 370 processor PC cluster of E\"otv\"os
University, on the 1024 processor PC cluster of Wuppertal University, 
on the 107 node PC cluster
equipped with Graphical Processing Units at Wuppertal University and on the
BlueGene/L at FZ J\"ulich. 
We used a modified version of the publicly available MILC code \cite{MilcCode}
with next-neighbor communication architecture for PC-clusters
\cite{Fodor:2002zi}.
 
\bibliographystyle{JHEP}
\bibliography{tc_8}

\newpage


\end{document}